\documentclass[10pt,aps,prd,twocolumn,showpacs,superscriptaddress,nofootinbib,nobibnotes,longbibliography,floatfix]{revtex4-1}

\usepackage[utf8]{inputenc}
\usepackage[T1]{fontenc}
\usepackage{comment}
\usepackage{bm}
\usepackage{mathtools,amsmath,amssymb,amsfonts,mathrsfs,eucal,graphicx,tensor,csquotes,accents,commath,chngcntr,siunitx}
\usepackage[dvipsnames]{xcolor}
\usepackage[unicode]{hyperref}
\hypersetup{colorlinks=true, citecolor=MidnightBlue,
            linkcolor=MidnightBlue, urlcolor=MidnightBlue, linktocpage=true}
\usepackage[normalem]{ulem}

\DeclareMathAlphabet{\mathpzc}{OT1}{pzc}{m}{it}
\usepackage{enumitem}
\usepackage{siunitx}
\DeclareSIUnit \parsec {pc}
\DeclareSIUnit \msun {M_\odot}

\usepackage{orcidlink}

\newcommand{\B}{{\mathcal{B}}}
\newcommand{\IITGn}{Indian Institute of Technology, Gandhinagar, Gujarat 382055, India.}

\begin{document}

\author{Rajes~Ghosh~\orcidlink{0000-0002-1264-938X}}
\email{rajes.ghosh@iitgn.ac.in}
\affiliation{\IITGn}

\author{Sreejith~Nair~\orcidlink{0000-0002-6428-5143}}
\email{sreejithnair@iitgn.ac.in}
\affiliation{\IITGn}

\author{Lalit~Pathak~\orcidlink{0000-0002-9523-7945}}
\email{lalit.pathak@iitgn.ac.in}
\affiliation{\IITGn}

\author{Sudipta~Sarkar}
\email{sudiptas@iitgn.ac.in}
\affiliation{\IITGn}

\author{Anand~S.~Sengupta~\orcidlink{0000-0002-3212-0475}\vspace*{7pt}}
\email{asengupta@iitgn.ac.in}
\affiliation{\IITGn}

\date{\today}

\title{Does the speed of gravitational waves depend on the source velocity?}

\begin{abstract}
The second postulate of special relativity states that the speed of light in vacuum is independent of the emitter's motion. The test of this postulate so far remains unexplored for gravitational radiation. We analyze data from the LIGO-Virgo detectors to test this postulate within the ambit of a model where the speed of the emitted GWs ($c'$) from a binary depends on a characteristic velocity $\tilde{v}$ proportional to that of the reduced one-body system as $c' = c + k\, \tilde{v}$, where $k$ is a constant. We have estimated the upper bound on the  90\% credible interval over $k$ to be ${k \leq 8.3 \times {10}^{-18}}$, which is several orders of magnitude more stringent compared to previous bounds obtained from electromagnetic observations. The Bayes' factor supports the second postulate with a strong evidence that the data is consistent with the null hypothesis $k = 0$, upholding the principle of relativity for gravitational interactions.
\end{abstract}

\maketitle

\section{Introduction}
Our current understanding of gravity is based upon Einstein's theory of general relativity (GR), which describes gravity as a manifestation of spacetime curvature produced in the presence of matter/energy. A crucial prediction of GR is the existence of gravitational radiation that propagates along the null direction of the background metric. In particular, GR predicts that gravitational waves~(GWs) must travel with the speed of light ($c$) in vacuum. However, any putative non-GR modifications might lead to deviations from this fact. Therefore, like all other important predictions of GR, it is necessary to test the validity of this assertion as well.

For this purpose, let us consider GW emission from one of the most prominent sources observed by LIGO-Virgo collaboration, namely the binary merger events. Our goal is to investigate whether the speed of emitted GWs is independent of the underlying binary dynamics, as predicted by GR. Then, in the spirit of numerous prior investigations that employ bottom-up phenomenological techniques to constraint deviations from GR, we shall use a model where the GW speed ($c'$) depends on a characteristic velocity $\tilde{v}$ of the binary dynamics as $c' = c + k\, \tilde{v}$. Here, for simplicity, we shall assume $\tilde{v}$ to be proportional to the orbital velocity ($v$) of the reduced one-body. Though similar to many phenomenological models, ours also may not have any derivation from the first principle; they have become quite ubiquitous and instrumental in contemporary GW phenomenology.

To test the prediction of GR, one may consider the speed of emitted GWs from a binary as, $c' = c + \sum_{i} k_{i}\, v_{i}$. Here, $v_{i}$'s denote all possible choices of velocities in the binary dynamics that might effect the GW speed. According to GR, all $k_i$'s must vanish identically. However, since we will work under the post-Newtonian (PN) approximation and expand various physical quantities in series of the velocity of the reduced body, our calculation would greatly simplify if the deviation term depends only on the velocity of the reduced body. Moreover, our model is a natural and simple in the sense that it introduces a minimum number of extra parameters (only $k$), whose value has to be estimated from GW observations. Interestingly, other choices of velocity such as the velocities of component masses do not change our result significantly. It is because the velocity of the reduced mass set the scale of velocity in a binary and especially, for a comparable mass system such as \texttt{GW170817} under consideration, the velocities of component masses are of same order of magnitude as that of the reduced mass.

A curious reader may notice an interesting similarity between our model for GWs with that of ``emission theories" for light, which encodes the deviation from the second postulate of relativity by invoking a source-dependent speed of light~\cite{Ritz, Brecher:1977dm}. Nevertheless, our aim is strikingly different as we want to investigate whether the speed of emitted GWs depends on the binary dynamics. Hence, one should not confuse the velocity $v$ in our model with that of the GWs' source velocity.

While we are at it, let us also point out that these emission theories for light obey the principle of relativity by having no preferred frame for light transmission but assert that light is emitted at speed $c$ relative \textit{only} to its source. In these models, electromagnetic radiation emitted from a source moving with relative velocity $v$ along the direction of a stationary observer would propagate with velocity  \mbox{$c' = c+k\, v$~\cite{Ritz, Brecher:1977dm},} where $k$ parameterizes the deviation from the second postulate of relativity. Over years, several experiments and observations have been used to bound this parameter. The initial such constraints came from the classic work by de~Sitter~\cite{deSitter1, deSitter2} based on the observation of a binary star system. Fox~\cite{Fox:67} later pointed out the limitations of the approach mentioned above, owing to the presence of interstellar matter~\cite{1962AmJPh..30..297F}. Then, experiments designed to measure the velocity of \mbox{$\gamma$-rays} from $\pi^0$ mesons provided a strong upper bound ${k < 10^{-4}}$~\cite{1964PhL....12..260A}. The observation of pulsating X-ray sources in binary star systems has provided the most stringent bound on $k$ so far, as stated in~\cite{Brecher:1977dm}. An upper bound ${k < 2 \times 10^{-9}}$ can be estimated from the arrival time of pulses emitted by these sources.

Such tests have remained unexplored for gravitational radiation. We bridge this gap by studying our model where the speed of the emitted GWs ($c'$) from a binary depends on velocity  $\tilde{v}$ proportional to that of the reduced one-body system ($v$) of the binary dynamics as $c' = c + k\, \tilde{v}$. More details of the model can be found in the following section. Then, to constrain the deviation from the prediction of GR, we analyse the second science run of the advanced LIGO and Virgo detectors containing the GW signal \texttt{GW170817} from a binary neutron star inspiral and find the upper bound on the symmetric $ 90\% $ credible interval (CI) to be ${k \leq 8.3 \times 10^{-18}}$. We also obtain a logarithmic Bayes factor $\log_{10} {\mathcal{B}}_{01} \simeq 1.82$, in favour of the data under the null-hypothesis ($k=0$), thereby upholding the prediction of GR that the speed of emitted gravitational radiation is independent of the binary dynamics.

Let us also emphasize that our work is distinct from previous studies that assume a frequency-dependent speed of GW, which is modelled either by Lorentz-violating dispersion relations~\cite{KOSTELECKY2016510, Mirshekari:2011yq}  or by positing the existence of massive gravitons~\cite{Will:1997bb}. Constraints on such dispersive propagation have been obtained using data from GW detectors~\cite{TGR_GWTC2, O3bTGR2021}. Constraints on Lorentz and CPT-violating effects during gravitational wave propagation have also been studied~\cite{SMEdispersion, PhysRevD.106.084005, Niu_2022} under the standard-model extension (SME) framework using 45 high-confidence events detected by the LIGO-Virgo network. In the case of Lorentz violation, a preferred frame is required to measure the GW's energy and momentum and specify the form of the modified dispersion relation, which violates the first postulate of relativity. In contrast, our analysis does not invoke any preferred frame for GW propagation. Similarly, the frequency-dependent speed of a massive graviton can be attributed to its non-zero mass. In contrast, in this work, the speed of GW is solely determined by the characteristic binary velocity $\tilde{v}$. In addition, this velocity is not be uniform and may even vary from one source to another. As such, our result should not be interpreted as a bound on the GW speed in contrast to previous research~\cite{TGR_GWTC2, LIGOScientific:2017zic}.
\section{Geometric setup}\label{setup}
\begin{figure}[t]
\centering
    \includegraphics[width=0.95\columnwidth]{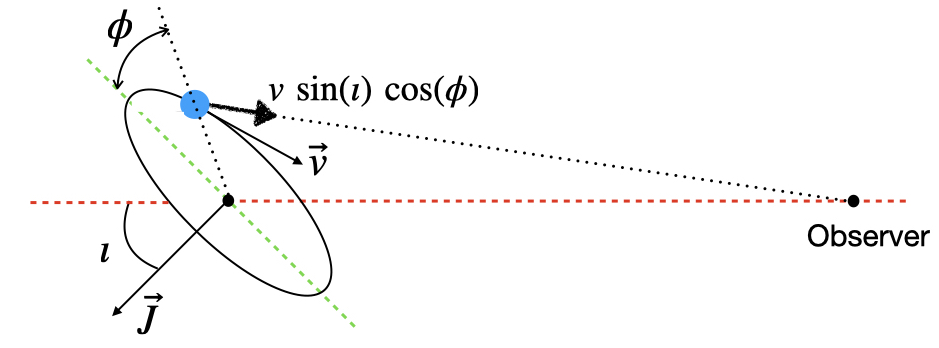}
    \caption{Geometric orientation of the reduced binary system relative to a distant observer (for \texttt{GW170817}, the estimated inclination angle of the effective one-body orbit is $\iota = 151^{\circ}$ using a post-Newtonian waveform model), $v$ is the orbital velocity, and $\vec{J}$ represents the direction of the angular momentum.}
\label{fig:orientation}
\end{figure}

It is useful to describe the geometric setup into two parts: the generation and propagation of GW, and discuss them separately. GW are emitted from a compact binary system with component masses $m_1$ and $m_2$ rotating around their common centre of mass. We shall treat the radius of the circular orbit to vary adiabatically due to an induced gravitation radiation reaction and neglect tidal effects for simplicity. Such a problem is well understood within GR using the framework of restricted post-Newtonian (PN) formalism, which expresses the amplitude $A(f)=\mathcal{A}\, f^{-7/6}$ through the quadrupole approximation and the phase $\psi(f)$ is given by an expansion in powers of ${\beta(f) := v(f)/c}$~\cite{Blanchet:2001aw, Blanchet:2013haa}: 
\begin{equation} \label{waveform}
    h(f) = \mathcal{A}\, f^{-7/6}\, e^{i\, \psi(f)}\, ,
\end{equation}
where $f$ is the frequency of the emitted GW and ${v = (\pi\, \text{G}\, m\, f)^{1/3}}$ is the orbital speed of the reduced one-body of mass $\mu = m_1\, m_2/ m$ with respect to the centre of mass,  ${m = m_1+m_2}$ being the total mass. The orientation of such an orbit is shown in Fig.~(\ref{fig:orientation}).

One should note that under the consideration of our model, the emitted GW has a frequency-dependent velocity, ${c'(f) = c + k\, v(f)\, \sin(\iota)\, \cos(\phi(f))}$ with respect to the distant observer, where the angular factors are due to the projections of the orbital velocity along the observer. This seems to suggest that we have used the projected component of the orbital velocity of the effective one-body system as a choice for velocity in our model, in the spirit of the emission theories of light. However, such a construct is not physically permissible, as in our case, we want every physical quantity to be invariant under the relabeling ($m_1\leftrightarrow m_2$) of the component masses. This is true for the magnitude of the relative velocity $v$ and the inclination angle $\iota$, but not true for the factor $\cos \phi$. Under such a relabeling, the factor $\cos \phi$ changes sign and thus, the model for the frequency-dependent velocity ${c'(f) = c + k\, v(f)\, \sin(\iota)\, \cos(\phi(f))}$ is not relabelling invariant. Therefore, we must prescribe a model that is manifestly relabelling invariant. This can be achieved by the following simple modifications of the previous model: 

\begin{enumerate}
\item $c'(f) = c + k\, v(f)\, \sin \iota \, \lvert \cos \phi \rvert$\, ,

\item $c'(f) = c + k\, v(f)$\, ,

\item $c'(f) = c +  \frac{k\, ( m_1 - m_2)}{m_1 + m_2}\, v(f)\, \sin \iota \, \cos \phi $ .
\end{enumerate}
The first model considers only the magnitude of the projection factor $\cos \phi$. The second model does not use the projection, such that the speed of gravity depends solely on the magnitude of the relative orbital velocity. Whereas the third model uses a mass-dependent factor to make the frequency-dependent part relabelling invariant. All these models are simple choices which are explicitly relabelling invariant introducing at most linear corrections to the speed to gravity proportional to the relative velocity \footnote{We sincerely thank the anonymous referee for pointing out the importance of relabeling invariance and suggesting the third model for the frequency-dependent GW velocity.}. We could have also included a relabelling invariant second-order term in the orbital relative velocity. But, such a term would have a negligible effect in our analysis. In the absence of a well-defined theory to generate such corrections, we interpret these as possible phenomenological models to study and verify the universality of the speed of GWs. We have performed our analysis with all these three models and find constraints on the parameter $k$. For simplicity and the presentation purpose, we will explicitly discuss the calculations for the first model alone and only quote the results for others.

The luminosity distance $d_L$ to the source is the largest length scale in the problem, overshadowing any other length scale (such as orbital size) associated with the GW generation process. As such, the effects of $k$ will be captured more prominently in the propagation of GW rather than in its generation, which we shall consider given by the PN analysis. Also, since the corresponding redshift of the event \texttt{GW170817} is ${z =0.0099 \ll 1}$,  we consider the propagation of GW in a Minkowski background. However, one can easily extend our analysis for sources at large $z$.

\section{Modelling the effect of emitter's speed on GW signals}
We shall assume the generation of the GW waveform from the binary to be well approximated by the restricted PN formalism within GR. We should emphasize again that this is a consistent approach since any generation effect such as ${k\, \beta^{2n}}$ (with ${\beta = v/c}$) at higher PN order $n$ will be completely overshadowed by the propagation effects proportional to $k\, d_L$. To model the effect of the dependence of the speed of GW on the emitter's speed, we shall use the first model discussed in the previous section, such that the gravitational waves propagate at speed, ${c'(f) = c + k\, v(f)\, \sin(\iota)\, \lvert \cos \phi(f) \rvert}$ as measured by the observer.

Then, for a body (emitter) moving on a quasi-circular orbit with some inclination angle $\iota$ and at a distance $d_L$ from the observer, its angular frequency at the time $t_e$ is given by $\omega_e = 2 \pi f_e$ and the radius of the orbit is $R_e \simeq v_e/\omega_e$. Assuming the emitter has a small peculiar velocity, we can obtain a relation between the time ($t_e$) of GW emission and time of receiving ($t_o$) at the observer's location as,
\begin{align}\label{to}
    t_o \simeq t_e + \frac{d_L}{c} &-\frac{R_e\, \sin(\iota)}{c}\sin(\omega_e t_e) \nonumber \\ 
    & -\frac{k\, d_L\, v_e\, \sin(\iota)}{c^2} \lvert \cos(\omega_e t_e) \rvert\, . 
\end{align}
Now, let us consider the emission of two successive GW signals at source times $t_e'$ and $t_e = t_e' +\Delta t_e > t_e'$. Here, we are using the shorthand $x(t_e) = x_e$ and $x(t'_e) = x'_e$ for any quantity $x$ under consideration. Due to the frequency-dependent speed of GW propagation, these signals will reach the distant observer within a duration given by $\Delta t_o \neq \Delta t_e$, which can be calculated from Eq.~\eqref{to}. Then, with this set up, the GW phasing can be expressed as the following integral,
\begin{equation}
    \psi(f_e)=2 \pi \int_{f_e'}^{f_e}\left(t-t_c   \right) \, df + 2 \pi f_e t_c -\phi_c - \frac{\pi}{4}\, .
\end{equation}
We replace $t-t_c=\Delta t_o$ using $\Delta t_e$ obtained from Eq.~(\ref{to}) and use the radiation-reaction equation~\cite{Cutler:1994ys, Will:1997bb} to perform the integral over frequency. Absorbing all integration constants into $\tilde{t}_c$ and $\tilde{\phi}_c$ we get the phase as
\begin{widetext}
\begin{equation}
\label{phase}
\begin{aligned}
\psi(f_e) = \ & 2 \pi f_e \tilde{t}_c - \tilde{\phi}_c - \frac{\pi}{4} + \sum_{n}a_n \, \beta^{2n} \\
          & -3\epsilon_1 \left(\frac{v_e\, \sin(\iota)}{\sqrt{2}\, \pi\, c}\right) \Bigg[-   
            x^{2} +\frac{x^{6}}{20} -\frac{x^{10}}{72}+\frac{5\, x^{14}}{832}-\ldots\Bigg] \\
          & +3\epsilon_2 \left(\frac{k\, d_L\, v_e^4\, \sin(\iota)}{\sqrt{2}\, \pi\, c^2\, G \, m}\right)\, \Bigg[1+x^{2} + \frac{x^{4}}{2} \log (x) -\frac{x^6}{8} 
            +\frac{5\, x^{8}}{128}-\ldots \Bigg],
 \end{aligned}
\end{equation}
\end{widetext}
where $a_n$ is the standard PN coefficient of order n, ${x={v'}_e/v_e < 1}$, and
${(\epsilon_1,\, \epsilon_2)}$ are two dimensionless order-unity numbers whose values depend on ${(f_e',\, f_e)}$. We have checked that the $\epsilon$'s are order unity positive numbers within the range of frequencies ${(f_e',\, f_e)}$ considered by us. Although one could numerically estimate exact values of $\epsilon$, we shall skip this as it will not change the order of magnitude of the estimate we get on $k$.
It should also be noted that though we are using the suffix `$e$' to represent quantities measured in the emitter frame, associated detector-frame quantities are the same with a high degree of accuracy. It is because the source (\texttt{GW170817}) is situated at a very low redshift. 

\begin{figure*}[t]
    \centering
    \includegraphics[width=0.95\textwidth]{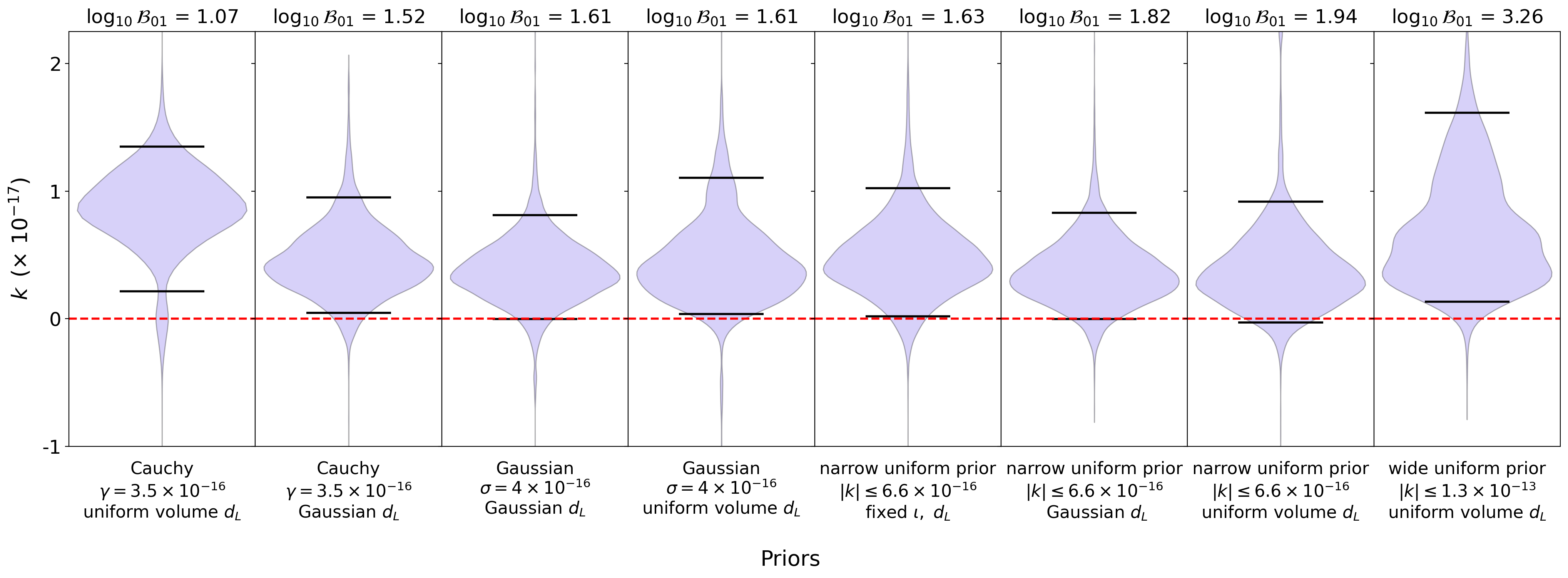}
    \caption{
    Marginalized posterior distribution ${p(k \mid \vec d)}$ are represented as violin plots for different prior configurations over $k$ (Model 1) and $d_L$  as labeled at the bottom of each panel. The corresponding logarithmic Bayes factors are indicated at the top. The horizontal red dashed line represents the second postulate of relativity $k=0$. In the fourth panel from the right, ($\iota = 151^\circ$, $d_L = 40 \, \si{\mega\parsec}$) are held fixed. The wide, non-informative prior (extreme right) gives the strongest evidence in favour of the null hypothesis, as expected (see the text for a discussion). Priors over other model parameters are described in the text and identical for each panel.}
    \label{fig:results_k}
\end{figure*}

\section{Measuring $k$ from LIGO-Virgo data}
We solve the stochastic inverse problem to estimate the numerical value of the parameter $k$ for the  first model and other parameters using \si{\num{360}~\second} of open archival data~\cite{GW170817data, RICHABBOTT2021100658} from the second observation run of the advanced LIGO and Virgo detectors containing the gravitational wave signal \texttt{GW170817}~\cite{PhysRevLett.119.161101} emitted by a merging binary neutron star system. We use the ``cleaned" data,  sampled at $ \si{\num{4096}\,\Hz}$, after a glitch was removed from the data collected at the LIGO-Hanford detector.

The posterior distribution ${p(\vec \theta \mid  \vec d ) \propto p( \vec d \mid \vec \theta ) \, p(\vec \theta)}$ 
over the eight-dimensional parameter space ${\vec \theta \equiv (k, \mathcal{M}, q, \chi_{1,2}, t_c, \iota, d_L)}$ is sampled using the dynamic nested sampling algorithm~\cite{10.1214/06-BA127, Higson_2018} implemented in the {\texttt{dynesty}} software package~\cite{Speagle_2020, sergey_koposov_2023_7600689}. 
Further, the phase-marginalised likelihood ${p(\vec d \mid \vec \theta)}$ is calculated using the \texttt{PyCBC} inference~\cite{Biwer2019} software package~\cite{alex_nitz_2022_6912865}. 
Here, $\mathcal{M}=\mu^{3/5}m^{2/5}$ and $q$ are, respectively, the chirp-mass and mass ratio of the compact binary system in the detector frame, $\chi_{1,2}$ are the dimensionless spins of the component masses, $t_c$ is the fiducial ``geocentric" coalescence time, $\iota$ is the inclination angle, and $d_L$ is the luminosity distance. 

The discovery of the optical counterpart and identification of the host galaxy NGC4993~\cite{Soares-Santos_2017} of this GW event allows us to uniquely fix other extrinsic parameters, namely the sky position (right-ascension $\alpha= \si{\num{3.44616}\,\radian}$ and declination $\delta= \si{\num{-0.408084}\,\radian}$). 
The polarization angle cannot be well estimated using data from a three-detector network using the PN signal model (flat posterior distributions indicating a lack of information): as such, this analysis uses a fixed polarization angle $\psi = 0$.\footnote{{We have also repeated our analysis with a standard uniform prior over $\psi$ between $(0, 2\pi)$ and found no significant change in the order of magnitude estimate of the parameters. See Appendix-\ref{app} for the details.}} As mentioned earlier, the coalescence phase is marginalized over when calculating the likelihood. 
The GW signal from the quasi-circular inspiral of the \texttt{GW170817} binary neutron star system is assumed to have an overall structure that closely resembles the restricted PN waveform model (given by Eq.~\eqref{waveform}) in the frequency domain; except that the phase-evolution is given by Eq.~\eqref{phase} containing modifications from the usual \mbox{3.5~PN} counterpart by the addition of extra terms due to the possible dependence of GW speed on the emitter's velocity through the proportionality factor $k$. As per the notation followed in this paper, ${k=0}$  corresponds to the case where the second postulate of SR holds. Thus, by measuring $k$, we can gauge the extent of deviation from this postulate as inferred from GW data.

A uniform prior distribution is assumed on the component masses and spins and over the coalescence time parameter around the event's coalescence epoch (GPS time ${\si{\num{1187008882.4}}}$). Further, we assume priors that are uniform in $(\cos\iota)$ for the inclination angle and uniform in co-moving volume for the luminosity distance $d_L$. For determining the prior range over $k$, the match ${m(k) = \left \langle h(k=0) \mid h(k) \right \rangle}$ between the standard \mbox{restricted 3.5~PN} waveform model (${k=0}$) against  waveforms that have non-zero values ${k\neq 0}$ was at first calculated. Overlap integrals were integrated up to a fiducial cut-off frequency $1593.18$ Hz\footnote{\label{footnote1}The frequency ${f_{\text{ISCO}} = 1593.18\, \si{\Hz}}$ corresponds to the innermost stable circular orbit of BNS system with component masses \mbox{$(1.41, 1.35) \,  \textup{M}_\odot$}  as determined from an earlier  analysis~\cite{PhysRevX.9.011001} of the GW170817 event carried out by the LIGO and Virgo collaboration.}. In this way, we set a wide, non-informative uniform prior over the range ${|k| \leq 10^{-13}}$, corresponding to a regime where ${m(k) \geq 0.05}$. Other priors over $k$ and $d_L$ were also considered, as explained later.

It is well known that the choice of priors affects the outcome of Bayesian analyses~\cite{Du2019}, and ``reasonable" prior distributions based on practical considerations need to be chosen to interpret the results accurately. For example, ultra-wide, non-informative priors are known to favour the null hypothesis~\cite {Rouder2009, Du2019}, which could bias the interpretation of the Bayes factor $\B_{01}$. As such, we present our Bayesian analysis for different choices of the prior distribution over $k$ as shown in Fig.~(\ref{fig:results_k}) including wide and narrow uniform priors, weakly informative prior distributions modeled by the Cauchy distributions~\cite{CauchyDistribution} with different scale factors ($\gamma$) and zero-mean normal (Gaussian) distributions with different standard-deviations ($\sigma$). Simultaneously, we also consider different priors over the luminosity distance: other than the uniform prior over co-moving volume, and we also consider a Gaussian prior~\cite{10.1093/mnras/staa2120} over this parameter inspired by a high-precision measurement of the luminosity distance to the host galaxy NGC4993 using electromagnetic observations~\cite{Cantiello_2018}. 

Fig.~(\ref{fig:results_k}) highlights the main result of this analysis, where the marginalized Bayesian posterior ${p(k \mid \vec d)}$  are plotted for different choices of the prior distributions.
In each case, we consider two models: the null hypothesis ${\mathcal{H}_0: k=0}$, and the alternative hypothesis ${\mathcal{H}_1: k\neq 0}$, in which $k$ is free to vary.
We observe that $\mathcal{H}_0$ is nested under $\mathcal{H}_1$, (i.e. $\mathcal{H}_0$ can be obtained from $\mathcal{H}_1$ by setting $k=0$).  As such, the Bayes factor $\B_{01}$ 
can be calculated by the Savage-Dickey density ratio~\cite{Dickey1971, WAGENMAKERS2010158}:
\begin{equation}
    \label{BayesFactor}
    {\mathcal{B}}_{01} = \frac{p(\vec d \mid \mathcal{H}_0)}{p(\vec d \mid \mathcal{H}_1)} = \frac{p(k=0 \mid \vec d, \mathcal{H}_1)}{p(k=0 \mid  \mathcal{H}_1)}.
\end{equation}
The seismic low-frequency cutoff is taken to be $\si{\num{20}\,\Hz}$ in all the cases shown in Fig.~(\ref{fig:results_k}). 

For a narrow uniform prior over ${|k| \leq 6.6 \times {10}^{-16}}$ (corresponding to $m(k) \geq 0.5$) and a Gaussian prior over $d_L$ with mean ${\langle d_L \rangle = 40.7}\,\si{\mega\parsec}$ and standard deviation ${\sigma_{d_L} = 3.3\,\si{\mega\parsec}}$, we measure the MAP value of ${k = 2.93\,^{+5.36}_{-2.95} \times 10^{-18}}$, with the point ${k=0}$ lying marginally within the \mbox{90\% CI} of the posterior distribution. The upper limit on the \mbox{90 \% CI} is found to be ${k \leq 8.3 \times {10}^{-18}}$, which indicates that any deviations from the second postulate are highly constrained. The smallness of this value is significant because it is several orders of magnitude smaller than previous bounds obtained from electromagnetic observations by Brecher \textit{et al.}~\cite{Brecher:1977dm}. 

\begin{table}[h!]
\renewcommand{\arraystretch}{1.5}
\begin{center}
\begin{tabular}{|c|c|c|c|}
\cline{2-4}
\multicolumn{1}{c|}{}
 & Model 1 & Model 2 & Model 3 \\ \hline
 Bounds on $k$ $(\times 10^{-18})$ & 8.30 & 8.22 & 144.99 \\ \hline
 Bayes' factors & 1.82 & 1.77 & 1.13 \\ \hline
\end{tabular}
\end{center}
\caption{Results for a narrow uniform prior over ${|k| \leq 6.6 \times {10}^{-16}}$, and a Gaussian prior over $d_L$ with mean ${\langle d_L \rangle = 40.7}\,\si{\mega\parsec}$ and standard deviation ${\sigma_{d_L} = 3.3\,\si{\mega\parsec}} $ for different models of the frequency dependent speed of GW described in section \ref{setup}.} \label{Table}
\end{table} 
We can perform a similar analysis on all the previously proposed models of the frequency-dependent speed of GWs. The calculations of the GW phasing and the corresponding data analysis can be performed straightforwardly. Here, we shall only quote the bounds on the parameter $k$ for all three models in Table~\ref{Table}.
\section{Discussions and Conclusion}
We also calculate the Bayes factor via Eq.~\eqref{BayesFactor}, which quantifies the relative support for the null hypothesis $\mathcal{H}_0$ (i.e., the second postulate holds) compared to the alternative hypothesis $\mathcal{H}_1$ (i.e., there is a deviation from the second postulate). The Bayes factor for this choice of priors is $\log_{10} \B_{01} = 1.82$, indicating ``very strong" evidence for $\mathcal{H}_0$ as per Jeffrey's interpretation~\cite{Jeffreys:1939xee}. 
We refer the reader to the accompanying Appendix-\ref{app} for an example of the marginalized posterior distributions and pairwise parameter correlations obtained for the \texttt{GW170817} event where all model parameters (excluding sky position) are allowed to vary. Moreover, as a consistency check, when the PE analysis was carried out by fixing $k=0$, the estimated values of all the parameters were nominally consistent with those obtained by an analysis carried out by the LIGO-Virgo collaboration~\cite{PhysRevX.9.011001} and those from a separate analysis~\cite{Bilby2020} using the BILBY~\cite{Ashton_2019} Bayesian inference library applied to the first LIGO–Virgo gravitational-wave transient catalog. 
After analyzing the data for different priors, $k=0$ is seen to lie within the 90\% CI of the marginal distribution ${p(k \mid \vec d)}$ for only a few selected prior options. This can be attributed to the combination of the waveform model and the information obtained from the data. Regardless of the prior distributions considered, we obtain numerically consistent upper limits for the parameter $k$ and the Bayes factors.

Let us also mention that future detectors like Laser Interferometer Space Antenna may be able to detect GWs even from extreme mass ratio inspirals~\cite{LISA:2017pwj}. Such binaries will spend considerably more time in the LISA sensitivity band, providing us with longer inspiral signals. This will, in turn, make the constraint on the departures from the second postulate of SR more stringent and open up new possibilities to eliminate other alternatives of SR and GR.\\
\begin{acknowledgments}
We thank Ian~Harry and Nathan~K.~Johnson-McDaniel for carefully reviewing the manuscript. We also thank K.~G.~Arun and N.~V.~Krishnendu for helpful comments. We also thank anonymous referee for her/his important comments and suggestions which have considerably improved the draft. R.~G. (PMRF ID: 1700531) and S.~N. (PMRF ID: 1701653) are supported by the Prime Minister's Research Fellowship, Government of India. L.~P. is supported by the Research Scholarship Program of Tata Consultancy Services (TCS). S.~S. acknowledges support from the Department of Science and Technology, Government of India under the SERB Core Research Grant Grant (CRG/2020/004562). A.~S. gratefully acknowledges the grant provided by the Department of Science and Technology, India, through the DST-ICPS (Interdisciplinary Cyber Physical Systems) cluster project funding. We thank the HPC support staff at IIT Gandhinagar for help and support. This research has made use of data or software obtained from the Gravitational Wave Open Science Center~\cite{gwosc_web}, a service of LIGO Laboratory, the LIGO Scientific Collaboration, the Virgo Collaboration, and KAGRA. LIGO Laboratory and Advanced LIGO are funded by the United States National Science Foundation (NSF) as well as the Science and Technology Facilities Council (STFC) of the United Kingdom, the Max-Planck-Society (MPS), and the State of Niedersachsen/Germany for support of the construction of Advanced LIGO and construction and operation of the GEO600 detector. Additional support for Advanced LIGO was provided by the Australian Research Council. Virgo is funded, through the European Gravitational Observatory (EGO), by the French Centre National de Recherche Scientifique (CNRS), the Italian Istituto Nazionale di Fisica Nucleare (INFN) and the Dutch Nikhef, with contributions by institutions from Belgium, Germany, Greece, Hungary, Ireland, Japan, Monaco, Poland, Portugal, Spain. KAGRA is supported by Ministry of Education, Culture, Sports, Science and Technology (MEXT), Japan Society for the Promotion of Science (JSPS) in Japan; National Research Foundation (NRF) and Ministry of Science and ICT (MSIT) in Korea; Academia Sinica (AS) and National Science and Technology Council (NSTC) in Taiwan.
\end{acknowledgments}

Data availability. Codes used in this analysis are publicly
available in a Github repository~\cite{Lpathak_github}.
\appendix

\section{Marginalised posterior distribution}\label{app}
\begin{figure*}[hbt]
    \centering
    \includegraphics[width=0.75\textwidth]{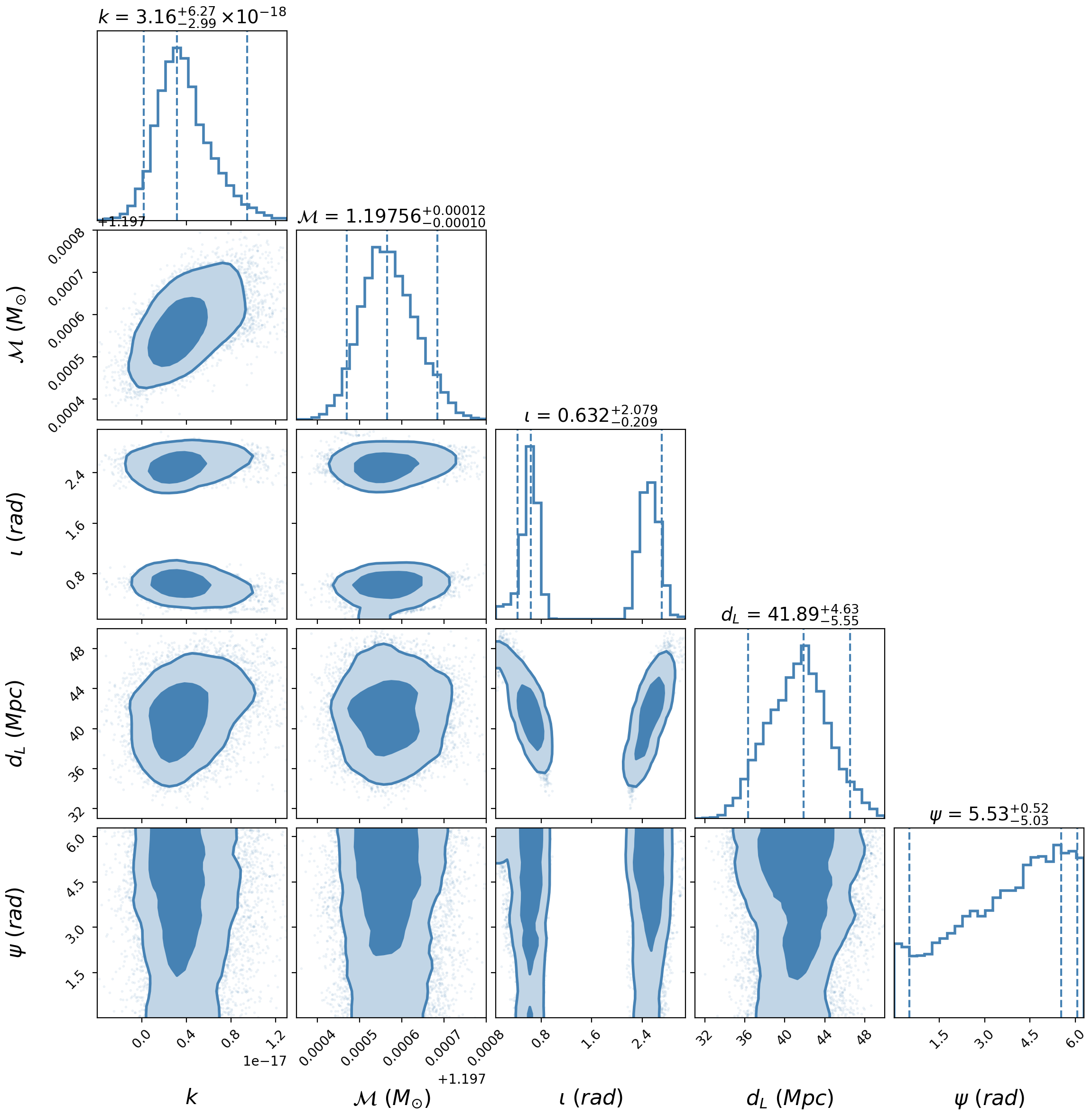}
    \caption{Marginalised posterior probability density over five (out of nine) model parameters, namely: $k$, $\mathcal{M}$, $\iota$, $d_L$ and $\psi$.
    A narrow uniform prior over ${|k| \leq 6.6  \times {10}^{-16}}$ and a Gaussian distribution over $d_L$ were used for this analysis. A uniform prior was taken over the polarization angle $\psi$. Priors over other model parameters are described in the text of the main paper. The polarization angle is poorly measured and is degenerate with all other parameters.}
    \label{fig:k-tc correlation}
\end{figure*}

In this appendix, we present the corner plot Fig.\,(\ref{fig:k-tc correlation}) which shows the pair-wise correlations between several parameters (including $k$ for model 1) and their marginalized posterior distributions (diagonal elements) for the analysis of the \texttt{GW170817} event where all model parameters (excluding sky position) were allowed to vary.

The data used in this analysis is from the LIGO and Virgo GW detectors containing the \texttt{GW170817} event. The waveform model (Eq.\eqref{phase} in the main text) used in this analysis allows for non-zero values of $k$, where the point $k=0$ corresponds to the second postulate of relativity. A narrow uniform prior over ${|k| \leq 6.6  \times {10}^{-16}}$ and a Gaussian distribution over $d_L$~\cite{10.1093/mnras/staa2120} over this parameter inspired by a high-precision measurement of the luminosity distance to the host galaxy NGC4993 using electromagnetic observations~\cite{Cantiello_2018} were used for this analysis. The prior distributions for the other model parameters are described in the main text.

For this choice of priors, the null $k=0$ value, corresponding to the second postulate, lies marginally outside the Bayesian 90\% credible interval. While the estimated chirp-mass agrees well with that estimated by the $k=0$ waveform model, inclusion of the parameter $k\neq 0$ in the expression for the phase evolution of the GW signal's dominant harmonic leads to a clear degeneracy between the inclination angle ($\iota$) and luminosity distance ($d_L$), as seen in the plot: we find that the inclination angle is split into a bimodal distribution for $k\neq 0$ models. Such degeneracy between these two parameters is commonly seen~\cite{GW190412} even for $k=0$ models that only consider the dominant harmonic of the GW signal. We observe that $\psi$ is poorly estimated from the data and is seen to be degenerate with all other parameters.

Similar corner plot analysis can be performed for the model 2 and model 3 also. The results are similar to that of model 1, with no appreciable modifications. 

The degeneracy between $\iota$ and $d_L$ parameters may be resolved after the inclusion of sub-dominant modes and/or the spin precession effects in the waveform model. Another possible extension of our work could be to consider more accurate waveform models containing the effects of tidal parameters and in-plane spins. We leave such extensions for future attempts.

%

\end{document}